\documentclass[mathleft
]{an}
\usepackage{graphicx}
\usepackage{times}
\begin{document}

\Pagespan{789}{}
\Yearpublication{2006}%
\Yearsubmission{2005}%
\Month{11}%
\Volume{999}%
\Issue{88}%

\title{
Fate of baby radio galaxies: Dead or Alive ?
}

\author{N. Kawakatu \inst{1}\fnmsep\thanks{Corresponding author:
  \email{kawakatu@th.nao.ac.jp}\newline}, 
H. Nagai \inst{1}
\and M. Kino \inst{2}
}
\titlerunning{Fate of baby radio galaxies}
\authorrunning{N. Kawakatu, H. Nagai, \& M. Kino}
\institute{
National Astronomical Observatory of Japan, 2-21-1, 
Osawa, Mitaka, Tokyo 181-8588, Japan
\and 
Institute of Space Astronautical Science,
JAXA, 3-1-1 Yoshinodai, Sagamihara 229-8510, Japan
}

\received{30 May 2005}
\accepted{11 Nov 2005}
\publonline{later}


\abstract{
In order to reveal the long-term evolution of relativistic jets 
in active galactic nuclei (AGNs), we examine the dynamical evolution 
of variously-sized radio galaxies [i.e., compact symmetric objects (CSOs), 
medium-size symmetric objects (MSOs), Fanaroff-Riley type I\hspace{-.1em}I 
radio galaxies (FRI\hspace{-.1em}Is)]. 
By comparing the observed relation between the hot spot size and 
the linear size of radio source with a coevolution model of hot spot 
and cocoon, we find that the advance speed of hot spots and lobes 
inevitably show the deceleration phase (CSO-MSO phase) and the 
acceleration phase (MSO-FRI\hspace{-.1em}I phase). 
The deceleration is caused by the growth of the cross-sectional area 
of the cocoon head. 
Moreover, by comparing the hot spot speed with the sound speed of 
the ambient medium, we predict that only CSOs whose initial advance speed 
is higher than 0.3-0.5c can evolve into FRI\hspace{-.1em}Is. 
}

\maketitle

\section{Introduction}
According to the standard model for radio-loud AGNs, 
a pair of relativistic jets transport away the bulk kinetic energy 
from the central compact 
region close to a central supermassive black hole 
(SMBH) to $l_{\rm h}\geq 10\, {\rm kpc}$ scale radio lobes, 
where $l_{\rm h}$ is the distance from the centre of the galaxies 
(Blandford \& Rees 1974). 
However, how young radio loud AGNs evolve into powerful extended 
radio sources (e.g., FRI\hspace{-.1em}I radio galaxies) 
is one of the primary problems in astrophysics. 
In order to clarify this issue, it is important to understand the 
nature of the small, young progenitors of FR I\hspace{-.1em}Is. 
Young and compact radio sources such as CSOs 
($l_{\rm h} <$ 1kpc) and MSOs ($l_{\rm h}=1-10$ kpc) have been discovered 
by many authors (e.g., Wikinson et al. 1994; Fanti et al. 1995; 
Readhead et al. 1996).
The bright CSOs and MSOs show the FRI\hspace{-.1em}I like 
morphology and possess hot spots (reverse shock) 
which are the signatures of supersonic expansions. 
The recent observations of several CSOs and MSOs have indicated that 
their age ($\sim 10^{2-5}\,{\rm yr}$) is shorter than 
the typical age of FRI\hspace{-.1em}Is (e.g., Parma et al. 1999), 
which implies that CSOs and MSOs are possible candidates as the progenitors 
of FRI\hspace{-.1em}Is (e.g., Owsianik, Conway \& Polatidis 1998; Taylor et al. 2000; Giroletti et al. 2003; Polatidis \& Conway 2003; Nagai et al. 2006). 

A number of authors have investigated the long-term evolution of extragalactic 
radio sources in different ways (e.g., Begelman \& Cioffi 1989: hereafter BC89; Fanti et al. 1995; Begelman 1996; Readhead et al. 1996; Kaiser \& Alexander 1997; Snellen, Schilizzi \& van Langevelde 2000; Perucho \& Mart{\'{\i}} 2002; 
Kawakatu \& Kino 2006, hereafter KK06). 
A constant advance speed of hot spots, or a constant aspect 
ratio of the cocoon (i.e., a self-similar evolution) have often 
been assumed for a dynamical evolution of radio-loud AGNs for simplicity. 
{\it Do these assumptions reflect the actual evolution of radio sources ?} 
In order to answer this question, it is essential to build up 
an appropriate model of radio sources without assuming the 
constant aspect ratio and the constant advance speed of hot spots 
such as that presented by KK06.

Recent observations suggested that the power law index for the evolution 
of hot spot size changes at the transition between the interstellar medium 
and intergalactic medium, i.e., $l_{\rm h}\sim 1-10\,{\rm kpc}$ 
(Jeyakumar \& Saikia 2000: hereafter J00; Perucho \& Mart{\'{\i}} 2003: 
hereafter PM03). 
Since hotspots are identified as the reverse-shocked region of 
the decelerating jet, the evolution of hot spot size could reflect 
the dynamical growth of radio sources of various scales. 
However, it was hard to derive the dynamical evolution of radio 
sources from previous work (J00 and PM03) because of lack of spatial 
resolution, the observational bias and small sample of radio sources. 
In this study, we first compile sample of CSOs, MSOs and FRI\hspace{-.1em}Is 
sources larger than in previous works, by considering the observational bias 
and being careful about the data quality. 
Then, from the direct comparison with KK06, we make clear a long-term 
evolution of advance speed of hot spots 
(Kawakatu, Nagai \& Kino 2008 for details: hereafter KNK08). 

\begin{figure}
\includegraphics[width=80mm,height=80mm]{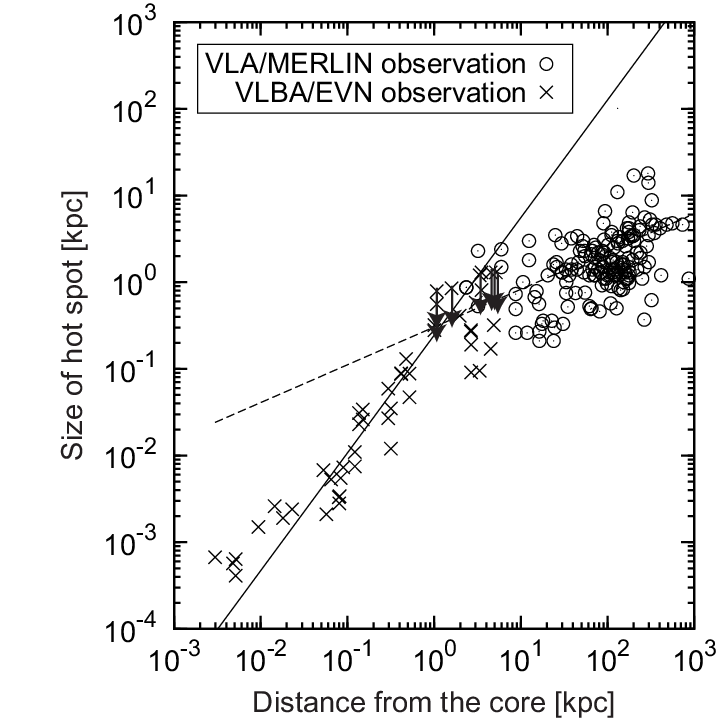}
\caption{
The relation of hot-spot size ($r_{\rm HS}$) and 
hot-spot distance from the core ($l_{\rm h}$).  
Crosses with arrows indicate upper limit.  
Solid line corresponds to the best-fit for the sources 
where $10^{-3}$~kpc$\le l_{\rm h}\le1$~kpc whereas broken line corresponds 
to that for the sources where $1$~kpc$\le l_{\rm h}\le10^{3}$~kpc. 
Note that upper-limit data were not included in the fittings. 
}
\label{label1}
\end{figure}

\section{Observed $r_{\rm HS}-l_{\rm h}$ relation}
Following the recent work of Nagai (2007), we compiled 117 variously-sized 
radio galaxies and examined the physical quantities of the hot spots. 
Based on these data, we focus on how the size of the hot spot changes with 
the distance from the core. 
Figure 1 shows the size of a hot spot, $r_{\rm HS}$ with respect to 
$\l_{\rm h}$, together with the best linear fit on the log-log plane 
($\log{r_{\rm HS}}=a\log{l_{\rm h}}+b$).  We estimated $r_{\rm HS}$ as $(\theta_{\rm maj}+\theta_{\rm min})/2$ where $\theta_{\rm maj}$ and $\theta_{\rm min}$ are full width at half maximum (FWHM) of major and minor axes of the component obtained by Gaussian model fittings, respectively.  The slope changes around $\sim1$~kpc, and then the best fit value of $a=1.34\pm0.24$ for $l_{\rm h} < 1\,{\rm kpc}$, while $a=0.44\pm 0.08$ for $l_{\rm h} > 1\,{\rm kpc}$. 
This tendency is consistent with PM03, but a large number of sources 
in our sample allow us to confirm the tendency more clearly. 
Concerning the possible uncertainties of the estimation of the hot-spot 
size and the slope change are discussed in KNK08 with more details. 
As a result, even if we allow for possible uncertainties, the trend of 
the observed $r_{\rm HS}-l_{\rm h}$ relation would not be changed. 
It seems reasonable to suppose that the slope change occurs around 1~kpc.

\section{Coevolution model of hot spot and cocoon} 
We briefly review here a dynamical evolution model of 
radio sources (KK06) which traces the dynamical evolution of 
advancing hot spots  and expanding cocoon (e.g., BC89; see also Kino \& 
Kawakatu 2005). 
We consider a pair of initially relativistic jets propagating in an ambient
medium with matter density ($\rho_{\rm a}$). Here $\rho_{\rm a}$ is given by 
$\rho_{\rm a}(l_{\rm h})\propto l_{\rm h}^{-\alpha}$. 
The basic equations of the cocoon expansion can be obtained as follows:
\begin{enumerate}
\item
The equation of motion along the jet axis, i.e., 
the momentum flux of a relativistic jet is balanced by the ram 
pressure of the ambient medium spread over the effective cross-sectional 
area of the cocoon head, that is, $A_{\rm h}$, 
$L_{\rm j}/c=\rho_{\rm a}v_{\rm HS}^{2}A_{\rm h}$, 
where $L_{\rm j}$, $c$ and $v_{\rm HS}$ are the 
total kinetic energy of jets, the light speed and the hot spot velocity, 
respectively. 
Here we assume that $L_{\rm j}$ is constant in time.

\item
The equation of motion perpendicular to the jet axis, 
that is the sideways expansion velocity, $v_{\rm c}$, which is equal to 
the shock speed driven by the overpressured cocoon with internal 
pressure ($P_{\rm c}$), 
$P_{\rm c}=\rho_{\rm a}v_{\rm c}^{2}$. 

\item 
The energy conservation in the cocoon, namely 
all of the kinetic energy transported by the jets 
is deposited as the cocoon's internal pressure, that is, 
$P_{\rm c}V_{\rm c}=2(\gamma_{\rm c}-1)L_{\rm j}t$
where $V_{\rm c}$ is the volume of the cocoon, $t$ is the 
life time of the source and $\gamma_{\rm c}=4/3$ is the specific 
heat ratio of the relativistic 
plasma in the cocoon. 
\end{enumerate}

The cross-sectional area of the cocoon body is given by 
$A_{\rm c}(t)=\pi l_{\rm c}^{2}\propto t^{X}$, 
where $l_{\rm c}=\int_{t_{\rm min}}^{t} v_{\rm c}(t') d t'$ 
is the radius of the cocoon body. 
Here $t_{\rm min}$ is the time at which the two-dimensional (2D) phase 
($A_{\rm h}$ growth phase) starts. 
Assuming that $A_{\rm h}/r_{\rm HS}^{2}$ is constant in time, 
$r_{\rm HS}$ and $v_{\rm HS}$ can be described in terms of the length 
$l_{\rm h}$. 
\begin{eqnarray}
r_{\rm HS}&\propto& {l_{\rm h}}^{S_{\rm r}}, \\
v_{\rm HS}&\propto& {l_{\rm h}}^{S_{\rm v}}. 
\end{eqnarray}
where $S_{\rm r}\equiv [X(-2+0.5\alpha)(\alpha-2)+3\alpha-4]/
[2X(-2+0.5\alpha)+6]$ and 
$S_{\rm v}\equiv [2-X(2-0.5\alpha)]/[X(-2+0.5\alpha)+3]$.
For a free parameter $X$, we can constrain the value of $1.2\le X \le 1.4$, 
by comparison with numerical simulations (Scheck et al. 2002; 
Perucho \& Mart{\'{\i}} 2007). 

\section{Evolution of the advance speed of hot spots}
\subsection{Constant velocity evolution ?}
In the current models of the evolution of powerful radio sources, 
the constant velocity of hot spots has often been assumed 
(e.g., Fanti et al. 1995; Begelman 1996; Readhead et al. 1996; 
Kaiser \& Alexander 1997). 
But, a deceleration of hot spots may take place via a strong 
interaction with denser ambient gas in host galaxies since 
CSOs and MSOs will have a significant interaction with the ambient medium 
as they propagate through it 
(e.g., Gelderman \& Whittle 1994; de Vries, Barthel \& O'Dea 1997
; Axon et al. 2000; Holt, Tadhunter, \& Morganti 2008; Labiano 2008). 
Thus, it is still unclear whether $v_{\rm HS}=const$ is a 
reasonable assumption on the evolution of radio sources. 
In order to test the validity of the $v_{\rm HS}=const.$ model, 
we will derive the required mass density profile to explain the observed 
$r_{\rm HS}-l_{\rm h}$ diagram. 
From eq. (2), the relation of $S_{\rm v}\propto 2-X(2-0.5\alpha)=0$ is needed 
to realize the constant velocity of hot spots. 
By eliminating $X$ in eq. (1), one finds 
$r_{\rm HS}\propto l_{\rm h}^{\alpha/2}$. 
Comparing this equation with the observed $r_{\rm HS}-l_{\rm h}$ relation, 
the constant velocity model requires that the slope of 
the inner part of the density profile ($< 1$ kpc) must be steeper than that of 
the outer part ($> 1$ kpc), i.e., $\alpha \sim 2.6\,(l_{\rm h} < 1\,{\rm kpc})$ and $\alpha \sim 0.8 \,(l_{\rm h} > 1\,{\rm kpc})$, as seen in Fig. 2 bottom 
(filled triangles and also the schematic picture). 
However, such a density profile of ambient matter is unrealistic 
because of the slope of the density profile 
in many clusters of galaxies and groups of galaxies ($l_{\rm h} > 1\,{\rm kpc}$) where $\alpha \approx 1.5$ (e.g., Trinchieri, Fabbiano \& Canizares 1986). 
Thus, the $v_{\rm HS}=const$ model can be rejected.

\begin{figure}
\includegraphics[width=80mm,height=100mm]{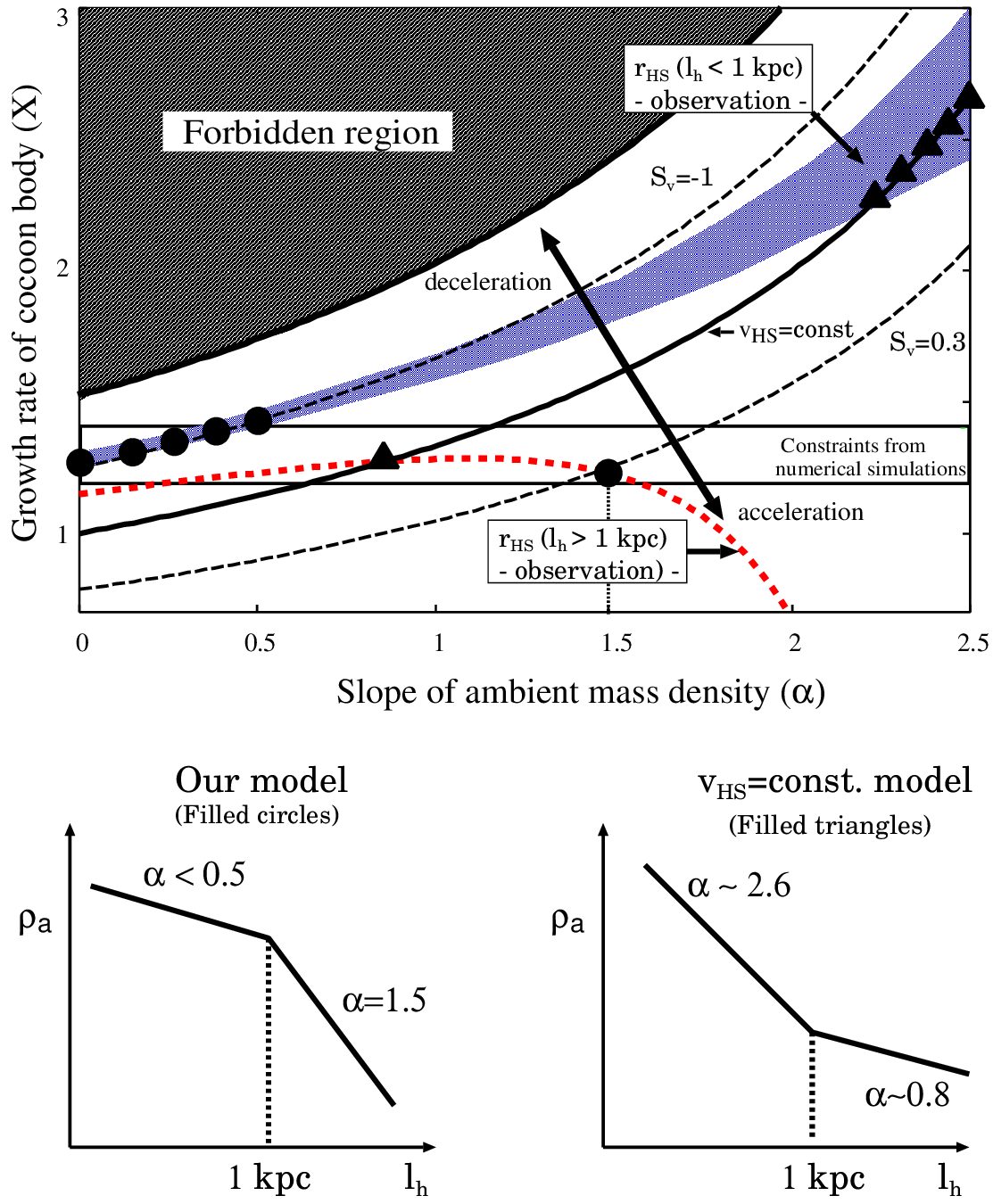}
\caption{
\small 
top) Growth rate of cocoon body ($A_{\rm c}\propto t^{X}$) 
against the value of exponents of the ambient matter density profile 
($\rho_{\rm a}\propto l_{\rm h}^{-\alpha}$). 
The dark grey (blue in the on-line version) region shows the allowed region 
given by the best fit value of the observed $r_{\rm HS}-l_{\rm h}$ relation for the CSO-MSO phase ($l_{\rm h} <$ 1 kpc), i.e., $S_{\rm r}=1.34\pm0.24$. 
The thick dotted line (red thick dotted line in the on-line version) 
corresponds to the allowed region obtained 
for the MSO-FRI\hspace{-.1em}I phase ($l_{\rm h} > $ 1 kpc), 
including the fitting errors, i.e., $S_{\rm r}=0.44\pm0.08$. 
The black solid line shows $v_{\rm HS}$=const ($S_{\rm v}=0$). 
The two black dashed lines are relations for the evolution of 
hot spot velocity $v_{\rm HS}\propto l_{\rm h}^{S_{\rm v}}$. 
The filled circles show the solutions for our model. 
We can constrain the power-law index $\alpha (l_{\rm h} < 1\,{\rm kpc})$ 
as $0 \leq \alpha < 0.5$. The filled triangles are solutions for 
the $v_{\rm HS}=const.$. 
bottom) The predicted ambient mass density profiles are shown in the 
bottom panel.
}
\label{label1}
\end{figure}

\subsection{Deceleration and acceleration of hot spots velocity}
We show an evolutionary track of radio sources that is consistent with 
the observed evolution of $r_{\rm HS}$ (see Fig. 1). 
In groups of galaxies and clusters of galaxies ($l_{\rm h} > 1\,{\rm kpc}$), 
it is well established that the slope of the ambient matter's density profile 
is $\alpha \approx 1.5$ (e.g., Trinchieri et al.1986). 
We assume here $\alpha=1.5$ for $l_{\rm h} > 1\,{\rm kpc}$. 
Comparing the KK06 model with the observed $r_{\rm HS}-l_{\rm h}$ relation, 
we can determine 
(i) the evolution of the advance speed of hot spots and 
(ii) the slope of mass density distribution for $l_{\rm h} < 1\,{\rm kpc}$. 
The predicted evolution of hot spots shows that the advance speed of 
the spots and lobes shows the deceleration phase (CSO-MSO phase) and 
the acceleration phase (MSO-FRI\hspace{-.1em}I phase) as follows 
(see filled circles showing allowed solutions in Fig. 2): 
In the CSO-MSO phase ($l_{\rm h} < 1\,{\rm kpc}$), the hot spot 
decelerates as 
\begin{equation}
v_{\rm HS}\propto l_{\rm h}^{-1}.
\end{equation}
In the MSO-FRI\hspace{-.1em}I phase 
($l_{\rm h} > 1\,{\rm kpc}$), the hot spot slightly accelerates as 
\begin{equation}
v_{\rm HS}\propto l_{\rm h}^{0.3}.
\end{equation}

For the CSO-MSO phase ($l_{\rm h} < 1\,{\rm kpc}$), 
a flatter density profile ($0 < \alpha \leq 0.5$) is predicted 
in order to satisfy both the observed $r_{\rm HS}-l_{\rm h}$ 
relation and the constraints from numerical simulations. 
The mass density profile is quite similar to a King-profile, 
as indicated by X-ray observations of elliptical galaxies 
(e.g., Trinchieri et al. 1986). 
The predicted mass density distribution implies that a significant 
interaction of the AGN jets with the ambient medium can take place 
during the CSO-MSO phase, as compared with the MSO-FRI\hspace{-.1em}I phase. 
Then, the interaction between the jets and the ambient medium is stronger 
in the CSO-MSO phase. 
This leads to the larger velocity of sideways expansion in order to 
maintain the energy conservation in the cocoon. 
Thus, $A_{\rm h}$ (or $r_{\rm HS}$) grows faster in the early phase 
of evolution. 
From the equation of motion along the jet axis, 
it is found that the advance speed of hot spots ($v_{\rm HS}$) 
is determined by the linear density of the effective working 
surface, $\rho_{\rm a}A_{\rm h}$, i.e., 
$v_{\rm HS}^{2}\propto (\rho_{\rm a}A_{\rm h})^{-1}$. 
When $\rho_{\rm a}A_{\rm h}$ increases with $l_{\rm h}$ in 
the CSO-MSO phase ($\rho_{\rm a}A_{\rm h}\propto l_{\rm h}^{2}$), 
the hotspot velocity decelerates.
On the other hand, the advance speed increases 
if $\rho_{\rm a}A_{\rm h}$ decreases with $l_{\rm h}$ 
in the MSO-FRI\hspace{-.1em}I phase 
($\rho_{\rm a}A_{\rm h}\propto l_{\rm h}^{-0.6}$). 
Summing up, the deceleration and acceleration of hot spot velocity 
is caused by the deviation from the 
balance between the deceleration effect via the 
growth of cocoon head and the acceleration effect due to the decrease 
of the ambient mass density.

\begin{figure}
\includegraphics[width=80mm,height=60mm]{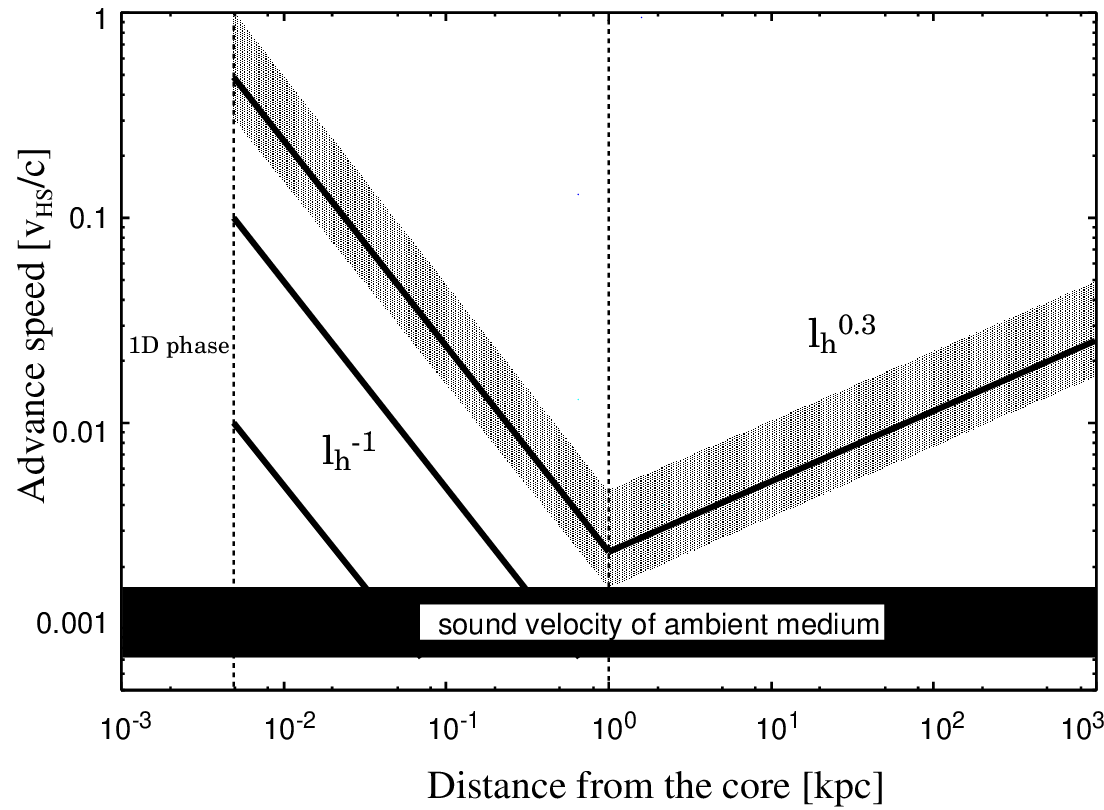}
\caption{
Our predictions of evolution of $v_{\rm HS}$, i.e., 
$v_{\rm HS}\propto l_{\rm h}^{-1}$ for $l_{\rm h} <\, 1{\rm \,kpc}$ 
and $v_{\rm HS}\propto l_{\rm h}^{0.3}$ for $l_{\rm h} >\, 1{\rm \,kpc}$. 
The black solid lines denote the evolution of $v_{\rm HS}$ 
for $v_{\rm HS}(l_{\rm h,2D})=0.01\, c, 0.1\, c,\,{\rm and}\,0.5\, c$ 
at $l_{\rm h, 2D}=5\, {\rm pc}$. 
The light grey shaded region represents the evolutionary path from 
CSOs into FRI\hspace{-.1em}Is.
The black shaded region shows the range of sound velocity of the 
ambient medium, i.e., $7\times 10^{-4}c < c_{\rm s} < 1.4\times 10^{-3} c$ 
($5\times 10^{6}\, {\rm K} < T_{\rm g} < 2\times 10^{7}\, {\rm K}$) 
where $T_{\rm g}$ is the temperature of the ambient medium.
}
\label{label1}
\end{figure}

\section{Prediction of fate of CSOs: Dead or Alive ?}
We further investigate which kind of CSOs can evolve 
into FRI\hspace{-.1em}I sources. 
For this aim, we compare the predicted evolution of $v_{\rm HS}$ with 
the sound velocity of the ambient medium, $c_{\rm s}$, 
since the supesonic jets can maintain at $l_{\rm h} > 10\, {\rm kpc}$ 
only when $v_{\rm HS} > c_{\rm s}$. 
For the slope of ambient matter density, 
we assume $\alpha \,(l_{\rm h} < 1\,{\rm kpc}) =0$ and 
$\alpha \,(l_{\rm h} > 1\,{\rm kpc}$)=1.5 (see $\S 4.2$). 
Correspondingly, the behavior of $v_{\rm HS}$ can be determined as
$v_{\rm HS}\propto l_{\rm h}^{-1}$ for $l_{\rm h} <\, 1{\rm \,kpc}$ 
and $v_{\rm HS}\propto l_{\rm h}^{0.3}$ for $l_{\rm h} >\, 1{\rm \,kpc}$ 
(see eqs. (3) and (4)). 
The hot ambient-gas temperature, $T_{\rm g}$ is  measured to be in the range 
of $T_{\rm g}=5\times 10^{6}\,{\rm K}-2\times 10^{7}\,{\rm K}$, 
i.e., $c_{\rm s}=(5kT_{\rm g}/3m_{\rm p})^{1/2}
\approx 7\times10^{-4}c-1.4\times10^{-3}c$ (e.g., Trinchieri et al. 1986), 
$k$ is the Boltzman constant and $m_{\rm p}$ is the proton mass. 

Figure 3 shows our predictions of the evolution of hot spot velocity 
for three initial advance speeds in the ambient medium with 
$v_{\rm HS}(l_{\rm h, 2D})=0.01c, 0.1c\, {\rm and}\, 0.5c$. 
Here we suppose $l_{\rm h, 2D}=5\,{\rm pc}$ 
where $l_{\rm h, 2D}\equiv v_{\rm HS}(l_{\rm h, 2D})t_{\rm min}$ 
is the distance from the core at which the 2-D phase starts. 
As seen in Fig. 3, we find that CSOs can evolve into FRI\hspace{-.1em}I 
sources, passing through MSOs for any $l_{\rm h}$ when $v_{\rm HS}(l_{\rm h, 2D})$ is larger than 0.3-0.5c because of $v_{\rm HS}(l_{\rm h}) > c_{\rm s}$. 
On the other hand, when $v_{\rm HS}(l_{\rm h, 2D})$ is less than 0.3-0.5c, 
$v_{\rm HS}$ is comparable to the sound velocity during the CSO-MSO phase 
($l_{\rm h} < 1\, {\rm kpc}$). 
Thus, the CSOs can evolve into FRI\hspace{-.1em}I sources, passing through 
distorted MSOs. 
Such distorted MSOs might correspond to low power compact 
radio sources of sizes of $\sim$ kpc (e.g., Kunert-Bajraszewska et al. 2005; 
Giroletti 2007). 

\section{Summary}
We have investigated the relation between CSOs, 
MSOs and FRI\hspace{-.1em}Is, by comparing the coevolution model 
of hot spots and a cocoon (KK06) with the observed $r_{\rm HS}-l_{\rm h}$ 
relation reflecting the dynamical evolution of radio sources. 
We find that the advance speed of hot spots and lobes strongly decelerate 
when the jets pass through the ambient medium in host galaxies 
(i.e., the CSO-MSO phase), while the jets accelerate 
outside host galaxies (i.e, the MSO-FRI\hspace{-.1em}I phase). 
The reason of deceleration is the growth of the cross-sectional area of 
cocoon head. The predicted deceleration of hot spots seems to be consistent 
with the recent observation show that the outflow velocity of the ionized gas 
around radio lobes may decelerate as $l_{\rm h}$ increaes (Labiano 2008). 
Furthermore, by comparing the hot spot speed with the sound speed of 
the ambient medium, we predict that only CSOs whose initial advance speed is 
higher than 0.3-0.5c can evolve into FRI\hspace{-.1em}Is. 
This indicates that the origin of the FRI/FRI\hspace{-.1em}I dichotomy 
could be related to the difference in the initial advance speed 
(Kawakatu, Kino \& Nagai 2008 in preparation).

\acknowledgements
We would like to thank A. Labiano, I. A. G. Snellen, M. Nakamura, 
S. Jeyakumar and T. Nagao for fruitful discussions. 
NK is supported by Grant-in-Aid for JSPS. 


\end{document}